\documentclass[manuscript]{aastex}

\def \msun   {\hbox{M$_\odot$}}
\def \rsun     {\hbox{R$_\odot$}}

\shorttitle{Circumstellar Disk in the $\eta$ Cha Cluster}
\shortauthors{Simon, Schlieder, Constantin,\& Silverstein}

\begin{document}

\title{{\it WISE} Detection of the Circumstellar Disk Associated with 2MASS J0820-8003 in the
  $\eta$ Cha Cluster}

\author{M. Simon\altaffilmark{1},  Joshua E.  Schlieder\altaffilmark{2}, 
Ana-Maria Constantin\altaffilmark{3}, and  Michele Silverstein\altaffilmark{4}}

\altaffiltext{1}{Department of Physics and Astronomy, Stony Brook University,
    Stony Brook, NY 11794-3800, USA  michal.simon@stonybrook.edu}
\altaffiltext{2}{Max Planck Institute f\"ur Astronomie, K\"onigstuhl,
  69117 Heidelberg, Germany}
\altaffiltext{3}{International Computer High School of Bucharest, Bucharest
030328,  Romania}
\altaffiltext{4}{Cornell University, Ithaca, NY 14850, USA}

\begin{abstract}

The Nearby Young Moving Groups (NYMGs) of stars are ideal for the
study of evolution circumstellar disks in which planets may form
because their ages range from a few Myr to $\sim 100$ Myr, about the
same as the interval over which planets are thought to form.  Their stars
are distributed over large regions of the sky.  Hence,   the  {\it
  Wide Field Infrared Survey Explorer (WISE)} which scanned the entire
sky in four bands from 3.4 to 22.1 $\mu$m provides a database
well-suited for the  study of members of the NYMGs, particularly those identified
after the eras of the {\it IRAS} and {\it Spitzer} observatories.
Here we report our study of the stars in the $\epsilon$ and  
$\eta$ Cha, TW Hya, $\beta$ Pic, Tuc-Hor, and AB Dor NYMGs.
The {\it WISE Preliminary Release Source Catalog}, which covers $57\%$
of the sky, contains data for $64\%$ of the stars in our search lists.
{\it WISE} detected the 11.6 and 22.1 $\mu$m emission
of all the previously known disks except for the coldest one, AU Mic.
{\it WISE} detected no disks in the Tuc-Hor and AB Dor groups, 
the two oldest  in our sample; the frequency of disks detected by {\it WISE}
decreases rapidly with age of the group. {\it WISE} detected a 
circumstellar disk associated with 2M J0820-8003, a pre-main sequence
star with episodic accretion in the $\sim 6$Myr old  $\eta$ Cha   cluster. 
The inner radius of the disk extends  close to the star, 
$\sim0.02$ AU and its luminosity  is about a tenth that of the star.
The episodic accretion is probably powered by the circumstellar
disk discussed here.

\end{abstract}

\keywords{stars: Pre-Main Sequence---- stars: Open Clusters and
  Associations: individual ($\eta$ Chamaeleontis)----planetary
  systems:protoplanetary disks----infrared:stars}

\section{Introduction}

Planet formation is thought to occur  5 to 100 Myr following the
birth of a star (Armitage 2010).  Stars in the Nearby  Young Moving 
Groups (NYMGs) span a similar range of ages, from the very 
young   $\eta$ Cha cluster at $\sim 6$ Myr,  
to the relatively old AB Dor group at $\sim 70$ Myr 
(Torres et al 2008, T08)\footnote{For consistency we use group
memberships, ages, and mean distances as given by T08 in their Table 2
and tables specific to the groups.  As  techniques improve
some of the memberships and values may change.  Therefore the values
cannot be accepted uncritically and the reader is referred to T08 for
a discussion of the issues.}.  This suggests
that the NYMGs are ideal for  a time-ordered study of the stages of
disk evolution and planet formation.   Indeed, astronomers already know that the 
$\beta$~Pic Moving Group, with stars at median distance 35 pc and 
age $\sim 10$ Myr (T08) hosts at least two debris disks, 
$\beta$~Pic's and AU Mic's, one exoplanet, $\beta$ ~Pic's,
and a brown dwarf orbiting PZ Tel  (Smith and Terrile, 1984;  
Lagrange et al. 2010; Liu et al. 2004; Biller et al. 2010).

Members of a given NYMG  are distributed over large regions of the
sky.  {\it The Wide-Field Infrared Survey Explorer} {\it(WISE)} Mission
surveyed the entire sky at  at 3.4, 4.6, 11.6, and $22.1 \mu$m  
(Wright et al. 2010); thus it provides an excellent database to 
search for emission of circumstellar disks that may be associated with
planet formation.  However, the  {\it WISE Preliminary Release Source 
Catalog}\footnote{http://wise2.ipac.caltech.edu/docs/release/prelim/}
(henceforth {\it WPRSC}) contains data for only $\sim 57\%$ of the sky 
(Wright et al. 2010).  Thus, a search of the catalog for members of the NYMGs  will 
find that data is not yet available for the entire sample.  
We selected   the $\eta$ Cha cluster and the $\epsilon$ Cha,  TW Hya, 
$\beta$ Pic, Tuc-Hor, and AB Dor Associations  for
our study because their ages span the interval believed to mark the
epoch of planet formation. We consider the $\eta$ Cha cluster and 
$\epsilon$ Cha Association together for convenience. Their 
location, ages, and kinematics are similar but distinct; T08 discuss
the issues involved in the possible connection of the two groups.

The study of the content, structure, and evolution of planet forming 
circumstellar disks is a very active field of research at the present
time.  Debris disks associated with stars in the
 NYMGs  have been identified by  {\it IRAS},  imaged with the {\it
 Hubble Space  Telescope}  and studied with the {\it Spitzer
 Observatory}.   Warmer circumstellar disks have
been studied from the ground by their near-infrared emission
and mm-wavelength molecular line emission.   Of the 
stars in our study and in the {\it WPRSC}, {\it WISE} detected all the 
previously known debris disks except one, AU Mic, in the $\beta$ Pic
Moving Group and identified one new  disk in the $\eta$ Cha
cluster.   We describe our results here. 

\section{Sample Selection and {\it WISE} Detections} 

To compile the source lists we used the membership lists of T08
in their entirety and supplemented them  with members proposed 
after the publication of T08's review.   Table 1 summarizes our
samples in each NYMG.  The ages and average distances (columns 
2 and 3) are from T08 (but, see footnote 1 and particularly \S3.2). 
Column 4 gives the numbers of members entering our search and 
column 5 gives the  number of detections in the  {\it WPRSC}.   
The ratio of the  {\it WISE} detections to stars searched is 0.64, 
close to the  value expected.  Column 6 provides references for 
the source lists.

The properties of  {\it WISE}  data affect our survey in 3 ways.  First,  the 
angular resolution is $\sim 6''$ at the shorter wavelengths and $\sim
12''$  at 22.1 $\mu$m.  Thus, {\it WISE} may not resolve small
separation  visual binaries among the NYMG members.  The flux 
of a NYMG member may also suffer confusion with a nearby unrelated star.  
Second,  the detectors enter a non-linear response regime for stars
brighter than $\sim$ 8.0, 6.7, 3.8, and -0.4 mag\footnote{{\it WISE} 
magnitudes are on the Vega scale.} at 3.4, 4.6, 12, and 22 $\mu$m, 
respectively\footnote {Explanatory Supplement, \S VI.4.d}.  The 
{\it Explanatory Supplement} indicates that for $6.0 <[3.4\mu m]< 8.0$ 
~mag the 3.4$\mu$m magnitudes\footnote{The notation [wavelength] indicates
a magnitude.}  are still reliable to better than 0.1 mag which is sufficient for 
this work.  No star in our sample was bright
enough at $11.6$ and $22.1 \mu$m to place its flux in the non-linear
response regime.  Third,  {\it WISE}  sensitivity limits were an issue 
only at $22.1 \mu$m at which the flux of some stars 
fell below the $5 \sigma$ detection threshold (Wright et al. 2010). 
We could not use these stars because our search for  circumstellar
material depends on the reliability of the $22.1\mu$m fluxes.

\section{Results}
\subsection{{\it WISE} color-color diagram}

Figure  1 is a color-color diagram of {\it WISE} magnitudes,
([3.4] - [4.6]) {\it vs} ([11.6] - [22.1]), for the NYMG members 
in our study and, for comparison,  main sequence 
(MS) stars without disks.   We selected MS stars 
 in the spectral type range 06V to M5V subject to the 
constraints that they have {\it WISE}  detections, are fainter 
than 6.0 mag at 3.4$\mu$m, and the SNR of their  22.1 $\mu$m 
detection exceeds 5.0.  The extinctions to these stars, as measured
by their $E_{B-V}$ color excess are small, the largest is $A_V \sim
0.8$ mag toward the O6V star.  Stars sufficiently hot  that  their long 
wavelength IR spectra are described by the  Rayleigh-Jeans 
approximation lie near the origin.  The cooler late K and M spectral 
type stars have slightly positive ([3.4] -[4.6]) color and ([11.6] -
[22.1]) color still close to zero.

In Fig. 1, HD 191089 and V4046 Sgr are in the $\beta$ Pic Moving Group
(BPMG).  HD 191089 is a classic debris disk (DD) star (Mannings and Barlow, 
1998, M08) and V 4046 Sgr is a T Tauri  binary with a circumbinary 
molecular disk (Kastner et al. 2008). That their disks are cold is
indicated by their red [11.6] - [22.1] color, but a [3.4]-[4.6] color 
indistinguishable from that of the MS stars.  Other well known DDs
such as $\beta$ Pic's are missing from Fig. 1 usually because
 they are too bright at $3.4 \mu$m, $[3.4]<6.0$ mag.  
This is the case for many disks discovered by {\it IRAS}. 
The debris disk of AU Mic, also in the BPMG,  does not show an
excess at any of the {\it WISE} wavelengths. Its debris disk, first 
detected by imaging its scattered light (Liu et al.,  2004)  is so cold that
its  emission in excess of photospheric appears finally 
at $70 \mu$m (Rebull et al., 2008).  Very cold debris disks such 
as AU Mic's escape  {\it WISE} detection.   

All the other stars in Fig. 1 lying away from the region populated 
by the MS stars are members of the $\epsilon$ Cha Association and
$\eta$  Cha Cluster.  Table 2 summarizes the {\it WISE} detections of
the members. Columns 1 and 2 give the stars' names and spectral types, 
columns 3, 4, 5, and 6  give their magnitudes at 3.4, 4.6, 11.6 
and 22.1 $\mu$m.   Column 7 lists the 22.1 $\mu$m
SNR, and columns 8 and 9 show the warning flags for the
3.4 and 22.1 $\mu$m magnitudes (too bright at 3.4 $\mu$m?, SNR too
low at  22 $\mu$m?). Column 9 provides a reference to the star's 
source, if not T08,  and prior designation of its associated disk,
DD indicating a debris disk, PDD a primordial/debris disk (eg. Rh07b), and
TT a T Tauri  disk. The ([3.4]-[4.6]) color of the stars plotted
 indicates that their disks are not as cold as, for example, HD 191089's. DZ Cha
and MP Mus belong to the $\epsilon$ Cha Association.  DZ Cha is associated
with a DD (Wahhaj et al. 2010, W10) and MP Mus with a 
PDD by Rhee et al (2007b; Rh07b).  These disks have been thoroughly
studied and we will not discuss them further. The
remaining 7 stars are in the $\eta$ Cha cluster; we will discuss
them in the following subsection.   Stars in $\epsilon$ Cha
Association and $\eta$ Cha cluster without color excesses in the {\it
  WISE} bands lie among the MS stars in Fig.1.

To identify the disks with 22.1 $\mu$m emission detected 
by {\it WISE} but that cannot be analyzed in a color-color diagram 
because their host stars are too bright at $3.4 \mu$m,  we plot 
in Fig. 2 the  ([11.6]-[22.1]) color of all the stars in our study
that are reported in the {\it WPRSC}.  Their color lies along 
the  ordinate.  The groups are separated along the abscissa.  Of the 
168 stars plotted, 19 have ([11.6]-[22.1])  color $\ge 1.0$, a color
distinct from that of MS stars (see Fig. 1) and indicates the presence 
of cool material\footnote{The  number 
plotted, 168, is less than the total number of stars detected, 192 
(see Table 1), because the others have  22.1 $\mu$m detections at 
SNR$< 5$}.    Table 3 lists the 19 stars.    {\it WISE} detected 
no stars with  ([11.6]-[22.1]) color  $\ge 1.0$ in the two oldest 
NYMGs, the Tuc-Hor and AB Dor Associations,  even though our 
input list for AB Dor was by far the largest.  The two stars closest
to the   ([11.6]-[22.1]) =1.0 cutoff are $\eta$ Cham in the $\eta$
Cha cluster and HD 164249 in the $\beta$ Pic Group with
([11.6]-[22.1])= 0.84 and 0.85, respectively; both are known
to be associated with debris disks (Rh07b; Zuckerman \& Song, 2004).

 Fig. 3 plots the ratio of the number having   ([11.6]-[22.1]) color$\ge 1.0$ to total 
number plotted in Fig. 2 {\it vs.} group age.  The fraction of 
stars hosting circumstellar disks decreases with age and none 
remain later than the $\sim 30$ Myr age of the Tuc-Hor 
Association.   The example of the AU Mic disk in the
$ \beta$ Pic moving group suggests that circumstellar disks start
evolve to colder disks 
at around 10 Myr age and that AU Mic's disk  is among the first 
in the BPMG to do so.   This result is entirely consistent with Rebull et al's (2008) 
finding from their analysis of the cold disks in the BPMG detected at 24 and 70 $\mu$m 
that the cold disk fraction decreases with age.

\subsection{ Disks in the $\eta$ Cha Cluster}

 The stars ET Cha, RECX 16, ES Cha, EN Cha, EP Cha, and EK Cha plotted in Fig. 1 
are members of the compact $\eta$ Cha cluster at distance $\sim 97$pc  identified by 
Mamajek et al. (1999).  2M J0820-8003, identified as an outlying member
by Murphy et al. (2010) is located $\sim 1.5^\circ$ distant from the 
center of the cluster.  All except for 2M J0820-8003
are X-ray sources.  The cluster has been studied extensively
from the ground (eg Luhman, 2008; Murphy et al. 2010) and the
{\it Spitzer Observatory}.  Megeath et al.  (2005) reported {\it IRAC}
photometry at 3.6, 4.5, 5.8, and 8.0 $\mu$m, Sicilia-Aguilar et al.
(2009) obtained IRS 7.5 to 30$\mu$m spectroscopy,  and Gautier et al. (2010)
presented {\it MIPS} photometry at 24, 70, and 160 $\mu$m.  
Sicilia-Aguilar et al. and Gautier et al. show spectral 
energy distributions  of the cluster members in Fig. 1,  except for 
2M J0820-8003, from the visible to {\it MIPS} wavelengths. 
For stars observed by {\it IRAC} and {\it MIPS}, {\it WISE} photometry
is completely consistent with theirs in regions of wavelength overlap.   
Interestingly, EK, EN, and EP Cha (RECX 5, 9,
and 11, respectively) show silicate emission and evidence for
significant grain processing from ISM size and structure 
(Sicilia-Aguilar et al, 2009).

Fig. 4 shows the SED of 2MASS J0820-8003 using its {\it WISE} fluxes
(Table 2) and 2MASS photometry at J, H, and K.  Murphy et al. (2010)
give its spectral type as M 4.3.  The solid line is the photospheric
spectrum of one of the main sequence M4 stars plotted in Fig. 1 normalized to
2M J0820-8003 at J.   The IR spectrum of 2M J0820-8003 
is similar to those of EN Cha and RECX 16  
(see Gautier et al 2010, Fig. 5) in that the excess emission associated 
with a disk is already evident at $3.4 \mu$m but the spectrum is 
not rising beyond $11\mu$m as steeply as that of EK Cha. 

\subsection{Properties of the Disk Associatd with 2MASS J0820-8003}

 We calculated the total disk luminosity of 2M J0820-8003,  $L_D$, 
relative to that of the star, $L_*$, by subtracting the spectrum 
of the stellar photosphere from the total spectrum in Fig. 4 and 
numerically integrating the difference. We found that $L_D/L_* \sim 0.12$,
about twice the values calculated similarly by Gautier et al (2010) for 
disk spectra in the $\eta$ Cha group to $70\mu$m.  It is however smaller
than the greatest fractional luminosity they reported, 0.19, for ET
Cha  (see Table 4).
Our value is a lower bound because the disk spectrum of 2M J0820-8003 
at wavelengths $> 22 \mu$m is unknown.  The broad   
IR spectrum of 2M J0820-8003 (Fig. 4) suggests that the disk temperatures
vary with distance from the star.  We fitted the disk spectrum
with two components radiating as black bodies.  We found that
$T_w = 825$K for a warm presumably inner disk component and $T_c = 220$K
for cooler, outer component provided the best fits (Fig. 4).  Integrating
the luminosities of the two components (we integrated the cold component 
to 100$\mu$m), we obtain $L_w/L_* = 0.058$ and $L_c/L_*= 0.041$ for 
the warm and cold components. It is difficult to decide whether their sum 
$(L_w+L_c)/L_*\sim 0.10$ or $L_D/L* \sim 0.12$ provides a more accurate 
measure of the fractional disk luminosity.  The former estimate is limited
by the unrealistic simplicity of a two-component model and the latter
by our ignorance of the spectrum beyond $22 \mu$m.   Regardless, the values
are relatively large compared to the fractional luminosities of other
members of the $\eta$ Cha group and also of debris disks in older NYMGs.
If the disk particles radiate as black bodies, those in the warm and
cold components lie at radial distances 0.022 AU and 0.082 AU, respectively,
corresponding to $\sim 5$ and 18 \rsun.   These radial distances are
lower bounds; if the particle emissivity were less than 1 the distances
would be greater.

\section{Discussion}

Murphy et al's (2011) spectroscopic observations of 2MASS J0820-8003
over 6-months showed that most of the time  its  line emission 
originates in the chromosphere and mass accretion at low level.
They also detected one instance of strong line emission attributable
to a hundred-fold increase in its accretion rate.  The circumstellar
disk detected by {\it WISE} extends close to the star (\S 3.3) and is
the probable the source of the episode of accretion.
Table 4 compares 2MASS J0820 to other
stars in the $\eta$~Cha cluster for which the fractional disk luminosity
$L_D/L_*$ (Col. 2) and  mass accretion rate, $\dot{M}$ (Col. 3), have 
been estimated.  We include ET Cha in Table 4 even though its accretion rate is
not known because its fractional disk luminosity is the highest of the 
stars observed by G08.  2MASS J0820 is 
noteworthy because  its variable accretion rate can be the largest of
the 4 stars and the optical depth of its disk, estimated by $\tau \sim
L_D/L_*$, is particularly large.  It appears that, of the stars in
Table 4, 2MASS J0820 is the last to start
clearing its circumstellar disk.   2MASS J0820 provides evidence 
that the initiation of disk clearing can spread over the $\sim 6$ Myr
estimated  age of the $\eta$ Cha cluster.  This implies that
the start of planet formation in the  group also spans a
comparable range of times.

\section{Summary}

We searched the {\it WISE Preliminary Release Source Catalog}
for evidence of circumstellar disks in the $\epsilon$ and 
$\eta$ Cha, TW Hya, $\beta$ Pic, Tuc-Hor, and AB Dor Nearby 
Young Moving Groups.   For the $\sim 64\%$ of group members 
included in the catalog, our results are:

\parindent=0.0in

1) {\it WISE} detected all the previously known disks except for the
coldest one, AU Mic.  Detection of cold disks such as AU Mic's
requires observations at $\lambda > 20 \mu$m.

2) Consistent with earlier findings, the fraction of stars with 
circumstellar disks decreases rapidly with age; we found 
none in the Tuc-Hor and AB Dor moving groups.

3) {\it WISE} detected a new circumstellar disk associated with 
2MASS J0820-8003 in the $\eta$ Cha group (M10; Murphy
et al. 2011).  The star and disk are noteworthy for their episodic
accretion, large disk luminosity relative to the star, and start of 
disk clearing delayed to $\sim 6$ Myr.

\vskip 1.0cm

We are grateful to Davy Kirkpatrick and Stan Metchev for advice at
an early stage of this project.  MS  thanks C. Chen, L. Ingleby,
J. Kastner, and S.T. Megeath for helpful conversations.
A-M. Constantin's participation was enabled by the support of the
Simons Foundation and M. Silverstein's contribution as an REU visitor
was supported by NSF Grant PHY-0851594. Our work was also supported 
in part by NSF Grant AST-09-08406.  We used data products from 
the {\it WISE} Mission and Two Micron All Sky Survey assembled 
and maintained at the Infrared Processing and 
Analysis Center/California Institute of Technology, funded by 
NASA and the NSF.  Our research has also used of the SIMBAD database 
operated at CDS, Strasbourg, France.

\centerline{\bf References}

 Biller, B.A. et al. 2010, ApJ, 720, L82

 Carpenter, J.M., Mamajek, E.E., Hillenbrand, L.A., and
  Meyer, M.R.  2009, ApJ, 705, 1646 (C09)

 Cieza, L.P. et al. 2011, ApJ, 741, L25  (C11)

  Gautier III, T.N., Rebull, L.M., Stapelfeldt, K.R, and
  Mainzer, A. 2008, ApJ, 683, 813 (G08)

 Ingleby, L. et al. 2011, ApJ, 743, 105 (In11)

  Kastner, H.H., Zuckerman, B., and Bessel, M 2008, A\&A,
  491, 829

  Kastner, J.H., Zuckerman, B., Hily-Blant, P.,\&
  Forveille, T. 2008, A\&A, 492, 469

 Kiss, L.L.,  et al., 2011, MNRAS, 411, 117  (K11)

 Lagrange, A.-M. et al. 2010, Science, 329, 57

 L\'epine, S. and Simon, M. 2009, AJ, 137, 3632 LS09)

 Looper, D.L., et al., 2010, AJ, 140, 1486 (L10)

 Liu, M. 2004, Science, 305, 1442

 Mamajek, E.E., Lawson, W.A., and Feigelson, E.D. 1999,
  ApJ, 516, L77

 Mannings, V. \& Barlow, M.J. 1998, MNRAS 497, 330 (M98)

 Megeath, S.T., Hartmann, L., Luhman, K.L. and Fazio,
  G.G. 2005, ApJ, 634, L113

 Murphy, S.J, Lawson, W.A. and Bessell, M.S., 2010, MNRAS,
  406, L50 (M10)

 Murphy, S.J, Lawson, W.A.,  Bessell, M.S., and Bayliss,
  D.R. 2011, MNRAS, 411, L51 

 Rebull,L.M. et al. 2008 ApJ, 681, 1484

 Rhee, J.H., Song, I.,  Zuckerman, B.,\& McElwain, M., 2007, ApJ,  660,
  1556 (Rh07a)

 Rhee, J.H., Song, I., \& Zuckerman, B., 2007, ApJ,  671,
  616 (Rh07b)

 Rodriguez, D.R., Bessell, M.S., Zuckerman, B., \&
  Koestner, J.H., 2011, ApJ, 727, 62 (R11)

 Schlieder,  J.E., L\'epine, S., \& Simon, M. 2010, AJ, 140, 119, 
(SLS10)

 Schlieder,  J.E., L\'epine, S., \& Simon, M. 2012, AJ, 143, 80 
(SLS12)

 Shkolnik, E.L. et al., 2011, ApJ, 727, 6  (S11)

 Sicilia-Aguilar, A., et al. 2009, ApJ, 701,1188

 Smith, B.A. and Terrile, R.J. 1984, 226, 1421

 Torres, C. A. O., Quast, G. R., 
  Melo, C. H. F., and Sterzik, M. 2008, in {\it Handbook of Star
    Forming Regions, Vol. II}, (Astron.Soc.Pacific:San Francisco),
  B. Reipurth, ed.  (T08)

 van Leeuwen, F. et al.  1997, A\&A, 323, L61

 Wahhaj, Z. et al. 2010, ApJ, 724, 835 (W10)

 Zuckerman, B. \& Song, I. 2004, ApJ. 603, 738

 Zuckerman, B. et al., 2011, ApJ, 732, 61

\clearpage

\begin{deluxetable}{lrrrrl}
\tabletypesize{\scriptsize}
\tablecaption{Selected Nearby Young Moving Groups\label{tbl-1}}
\tablewidth{0pt}
\tablehead{
\colhead{Group} & \colhead{Age (Myr)}  & \colhead{Mean Distance (pc)}&\colhead{Number input}
&\colhead{Number {\it WISE} Detections}&\colhead{Source List}
}
\startdata
$\epsilon$~and~$\eta$~ Cha &    6   &$108\pm9$& 38  & 36&T08, Tables 8 and 9\\
                                                &         &    &1    &   1&K11\\
                                                &         &    & 7   &6 &M10\\
                                                &         &    &       &&\\
TW Hya                                     &    8   &$48\pm13$& 22&  5&T08 \\
                                                     &         &&  2    & 0 &L10\\
                                                     &         &&  4    & 0 &R11\\
                                                    &         &&  2    & 0 &S11\\
                                                   &         &&       &&\\
$\beta$ Pic                                  &   10 &$31\pm21$& 50&41&T08 \\
                                                   &         &&   6 &  5&K11\\
                                                   &         &&  11 &  7& LS09, SLS10, SLS12\\
                                                   &         &&       &&\\
Tuc-Hor                                      &   30 &$48\pm7$& 44& 20&T08   \\
                                                    &        &&   2&  1&K11\\  
                                                   &         &&    8&  7 &Z11\\ 
                                                  &         &&       &&\\
AB Dor                                        &    70&$34\pm26$&   89  & 56&T08 \\
                                                   &         &&   7    &3&Z11\\
                                                   &         &&   9&4&SLS10, SLS12\\
                                                   &         &&     &  &\\
Total                                           &         && 302&192\\
\hline
\multicolumn{6}{l}{References: Kiss et al. 2011, K11; L\'epine and Simon
2009, LS09; Looper et al. 2010, L10;Murphy et al. 2010, M10}\\      
\multicolumn{6}{l}{Rodriguez et al. 2011, R11;  Schlieder, L\'epine, and Simon
2010, SLS10, 2011, SLS12; Shkolnik et al. 2011, S11}\\
\multicolumn{6}{l}{Torres et al 2008 (T08); Zuckerman et al. 2011, Z11}\\ 
\enddata
\end{deluxetable}

\begin{deluxetable}{llcccccccl}
\tabletypesize{\scriptsize}
\tablecaption{{\it WISE} Detections in the $\epsilon$ Cha Association 
and $\eta$ Cha Cluster\label{tbl-2}}
\tablewidth{0pt}
\tablehead{
\colhead{Star} & \colhead{SpTy}  &\colhead{$[3.4]$}&\colhead{$[4.6]$}&
\colhead{$[11.6]$} & \colhead{$[22.1]$}  & \colhead{ $[22.1]$~SNR}  &
\colhead{$[3.4] \le 6.0$}&\colhead{SNR $\le 5.0$}&{Comments, Refs.}\\
}
\startdata
 EG Cha & K4Ve      &  7.163  &   7.153   & 7.034  &6.981 &17.8&N&N&\\
$\eta$ Cha& B8V   &  5.746  &   5.606   &  5.269  &4.425 &42.3&Y&N&DD, Rh07b\\ 
  RS Cha & A7V+A8V  &  5.454  &   5.288   & 5.475  & 5.486 &28.4&Y&N&~\\ 
  EQ Cha& M3e       &  8.303  &   8.156   & 7.995  & 8.042 &  8.3 &N&N&~\\
HD 82879& F6V      &  7.793  &   7.814   & 7.755  & 7.704 &  8.5 &N&N&~\\
CP-68 1388&K1V   &  7.741& 7.765  & 7.656  & 7.607 &10.6  &N&N&~\\
 DZ Cha     & M0Ve &  8.209 & 7.674 & 4.507  & 1.861 & 81.2 &N&N&DD, W10\\
 T Cha       &  K0Ve  & 5.899   & 4.974 & 4.631  & 2.601 & 62.6&Y&N&TD, C11 \\
 GSC9415-2676&M3e& 9.364   & 9.218 & 9.052  & 8.660 &  4.1  &N&Y&~\\
 EE Cha     & A7V    & 6.109     &   6.089  & 6.108  & 6.094 & 27.2 &N&N&~\\
 $\epsilon$ Cha & B9V  &5.814      &   5.118  & 5.097  & 4.952 & 26.9 &Y&N&~\\
 HIP 58490&K4Ve  &    8.172  &   8.200  & 8.077   &7.947 &  8.0  &N&N&~\\
 DX Cha    &A8Ve   &   4.012   &   3.026  & 0.806&-0.707&116.0&Y&N&PMS, Rh07a\\ 
 HD 104467&G3V  &     6.825 &   6.814  & 6.751  &  6.727 & 17.3 &N&N&~\\ 
 GSC9420-0948&M0e&   8.112 &   8.045 &  7.884  &  7.815 &  9.2  &N&N&~\\
 GSC9416-1029&M2e&   8.746  &  8.592  & 8.445  &  8.649  & 4.1  &N&Y&~\\
HD 105923& G8   V &   7.080  &  7.087  & 7.022& 6.978 &18.0  &N&N&~\\ 
 GSC9239-1495&M0e&   8.759  &  8.678  & 8.514 &  8.347& 5.2  &N&N&~\\
 GSC9239-1572&K7e&   8.275  &  8.158  &  8.008  &   8.009 & 7.7  &N&N&~\\ 
  CD-74 712&K3e& 7.730  &  7.760  &  7.683  &   7.520 &10.8 &N&N&~\\
  CD-69 1055&K0Ve    &    7.447 &  7.475  &  7.372  & 7.241   &14.4 &N&N&\\
   MP Mus  & K1Ve &   6.592 &  6.185  &  4.031  &1.63      &63.9&N&N&PDD, Rh07b\\  
   RECX 18  &M5e  &   10.658 & 10.421 &  10.254&  9.357 &  0.5  &N&Y&~\\
   RECX 17  &M5e  & 9.975     &  9.912  &  9.744  &9.503   & 2.7 &N&Y&\\ 
   ES Cha    & M5e  & 10.727   & 10.356  &  8.968  &7.390   &12.7 &N&N&\\ 
   EH Cha   & M3e   &      9.345&  9.213   &  9.070   &8.939  &  4.0 &N&Y&\\ 
   HD 75505& A1V &   6.935&  6.959   &  6.950   &7.019   &19.0 &N&N&\\ 
    EI Cha    &M1e   &     8.534 &  8.462    & 8.292    &7.843   & 7.6&N&N&\\ 
    EK Cha   &M4e   & 9.721     &  9.483    & 7.835    &5.218   &35.2&N&N&PDD, Rh07b\\
    EL Cha   & M3e   &9.251      &  9.113    & 8.929    &8.837   &4.8   &N&Y&\\ 
    EM Cha  & K6e    & 7.532    &   7.534    & 7.399   & 7.423  & 8.2 &N&N&\\ 
    ET Cha   & M3e   & 8.563    &   7.849    &5.490    & 3.431   &47.2&N&N&\\
    RECX 16 & M5e   & 11.209  & 10.680    &9.018    &7.224    &16.5   &N&N&\\
    EN Cha   & M4e     &9.120     &  8.755     &7.210    &5.497&39.1  &N&N&PDD, Rh07b\\ 
    EO Cha   & M0e     &8.615     &  8.586     &8.430    &8.335    &6.4   &N&N&\\ 
    EP Cha    & K6e     &7.240     &   6.902    &5.362    &3.751   &51.9  &N&N&PDD,Rh07b\\ 
             &          &                &             &            &            &          &   &&\\               
RAVEJ1221-7116&K7   &8.200     &8.172        &8.035     &8.043        & 7.6&N&N&K11\\
              &          &                &             &            &            &          &   &&\\               
2MJ0801-8058   &M4.4&   10.096&      9.888&    9.625 &   9.023 &    3.4&N&Y&M10\\    
 2MJ0820-8003   &M4.3&      10.062&   9.646&    7.732 &  5.878  &  32.0&N&N&M10\\    
 2MJ0905-8134  &M4.9&        11.116&  10.859&  10.642&   9.284&     0.9&N&Y&M10\\    
 2MJ0913-7550  &M4.8&        10.823&  10.553&  10.381&   9.026 &    1.8&N&Y&M10\\     
 2MJ0955-7622  &M4.1&        10.288&  10.081&   9.864&   9.033 &    0.8 &N&Y&M10\\
 RXJ0902.9-7759 &M3 &         9.162&   9.038&   8.882&   9.001  &3.6 &N&Y&M10\\
\enddata
\end{deluxetable}

\begin{deluxetable}{lclcl}
\tabletypesize{\scriptsize}
\tablecaption{Stars with $[11.6] - [22.1] \ge 1.0$ mag \label{tbl-3}}
\tablewidth{0pt}
\tablehead{
\colhead{Group} & \colhead{Number}  & \colhead{Star}
&\colhead{$[11.6]-[22.1]$ mag}&\colhead{Comments}
}
\startdata
$\epsilon$~and~$\eta$~ Cha     &12 &$\eta$ Cha & 0.84& DD\\   
                                         &        & DZ Cha  & 2.64& DD\\
                                        &         & T Cha    &  2.03&TT\\
                                         &         & DX Cha & 1.51 &DD\\
                                        &          & MP Mus & 2.40&PDD\\
                                        &          & ES Cha  & 1.58&PDD\\
                                        &          & EK Cha & 2.62&PDD\\
                                        &          & ET Cha & 2.06&PDD\\
                                        &          & RECX 16&1.79&PDD\\
                                        &          & EN Cha & 1.71&PDD\\
                                        &          & EP  Cha & 1.61 &PDD\\
                                        &         &2M J0820-8003&1.85\\
                                                  &         &       &&\\
TW Hya                                        &    1  & TWA11A&3.89&HR 4796A, DD \\
                                                   &         &       &&\\
$\beta$ Pic                      &   7  & $\beta$ Pic     &2.43&DD \\
                                        &         &  V4046   Sgr   &2.98&DD\\
                                         &         & HD 17255    &1.20& DD\\
                                         &         &$\eta$ Tel    &1.41&DD\\
                                         &         & HD 164249 &0.85&DD\\ 
                                         &         & HD 181327 &1.88 &DD\\
                                         &         & HD 191089 &1.88 &DD\\
Tuc-Hor                                      &   0  &  &  &       \\
                                                  &         &       &&\\
AB Dor                                        &    0        &   & &    \\
\enddata
\end{deluxetable}

\begin{deluxetable}{llll}
\tabletypesize{\scriptsize}
\tablecaption{Comparison of T Tauris in the $\eta$~Cha Cluster \label{tbl-4}}
\tablewidth{0pt}
\tablehead{
\colhead{Star} & \colhead{$L_D/L_*$}  &\colhead{ $\dot{M}$}&\colhead{Refs}\\
 \colhead{}      & \colhead{}                 &\colhead{$10^{-11}\msun y^{-1}$} &\colhead{}\\
}
\startdata
2MASS J0820     & 0.10                          &2.5 - 620 & This work, M11\\
EK Cha (RECX 5)   &0.06                          &5              &G08  \\
EN Cha (RECX 9)   &0.04                          &4              &G08\\
EP Cha (RECX 11) & 0.04                          &4              &G08\\
                            &                                 &$\le30$    & In11\\
ET Cha (RECX 15) &0.19                          &                 &G08\\
\enddata
\end{deluxetable}

\clearpage
\begin{figure}
\begin{center}
\includegraphics[scale=0.8]{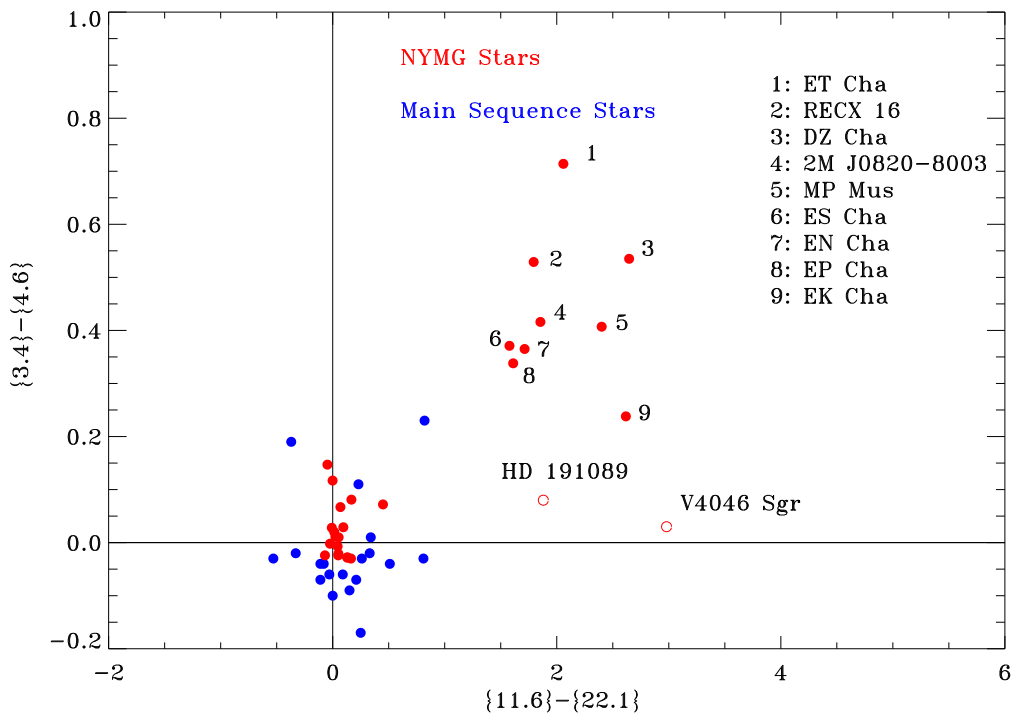}
\end{center}
\caption{The figure shows a $([3.4]-[4.6]) vs ([11.6]-[22.1])$
color-color diagram of the stars in our survey.  It includes only
stars with reliable 3.4 and 22.1
$\mu$m {\it WISE} magnitudes (see \S 3.1).  Filled red circles 
indicate {\it WISE} detections of  stars in the $\epsilon$ Cha Association and 
$\eta$ Cha Cluster and open circles indicate known debris disks
in the $\beta$ Pic group.  Disk identification in stars with reliable
$([11.6]-[22.1])$  colors are listed in Table 3.   The filled blue
symbols indicate  randomly selected main sequence stars of spectral
type O8V to M4V} 
\end{figure}

\clearpage
\begin{figure}
\begin{center}
\includegraphics[scale=0.8]{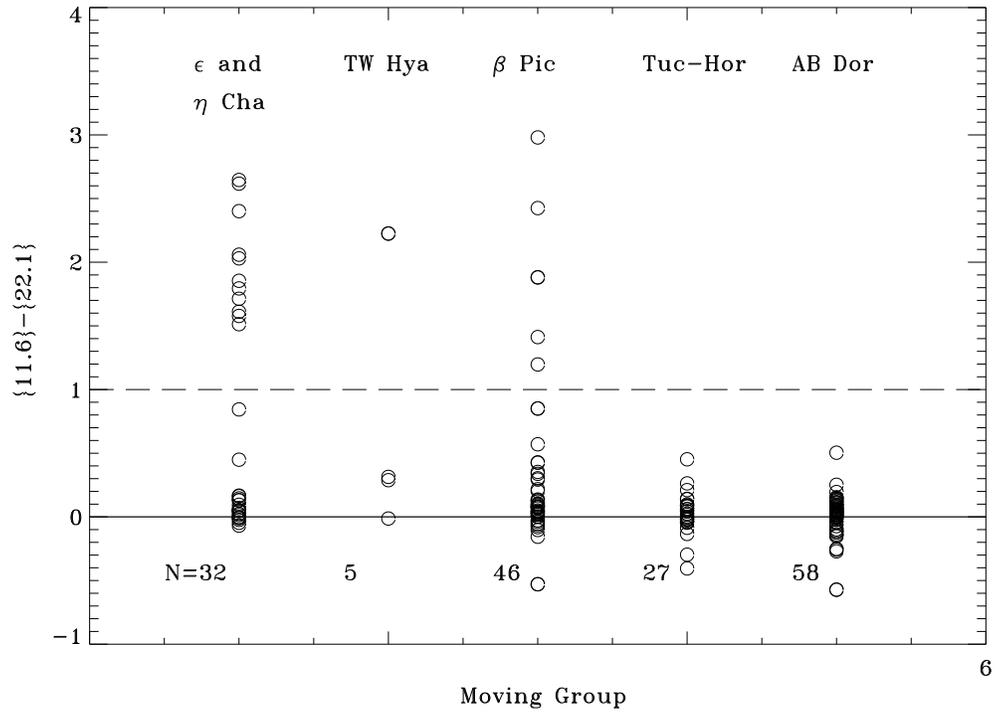}
\end{center}
\caption{Distribution of $[11.6] - [22.1]$ color for all the stars in
the groups we studied that are in the {\it WISE Preliminary Release
Catalog}.  $N$ gives the number by group, and the dashed line at
$[11.6] - [22.1]=1.0$ indicates an approximate threshold for
identification of stars with circumstellar  disks (see Fig. 1).}
\end{figure}

\clearpage
\begin{figure}
\begin{center}
\includegraphics[scale=0.8]{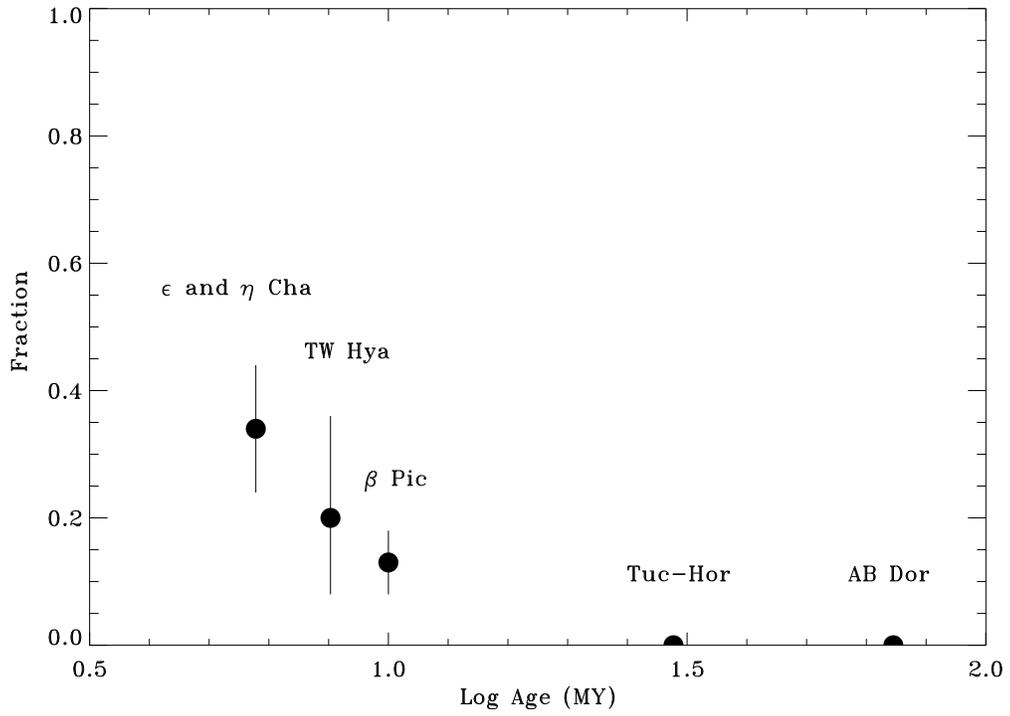}
\end{center}
\caption{Ratio of the number of stars with $[11.6] - [22.1]\ge 1.0$
to total number {\it vs} log group age.}
\end{figure}

\clearpage
\begin{figure}
\begin{center}
\includegraphics[scale=0.8]{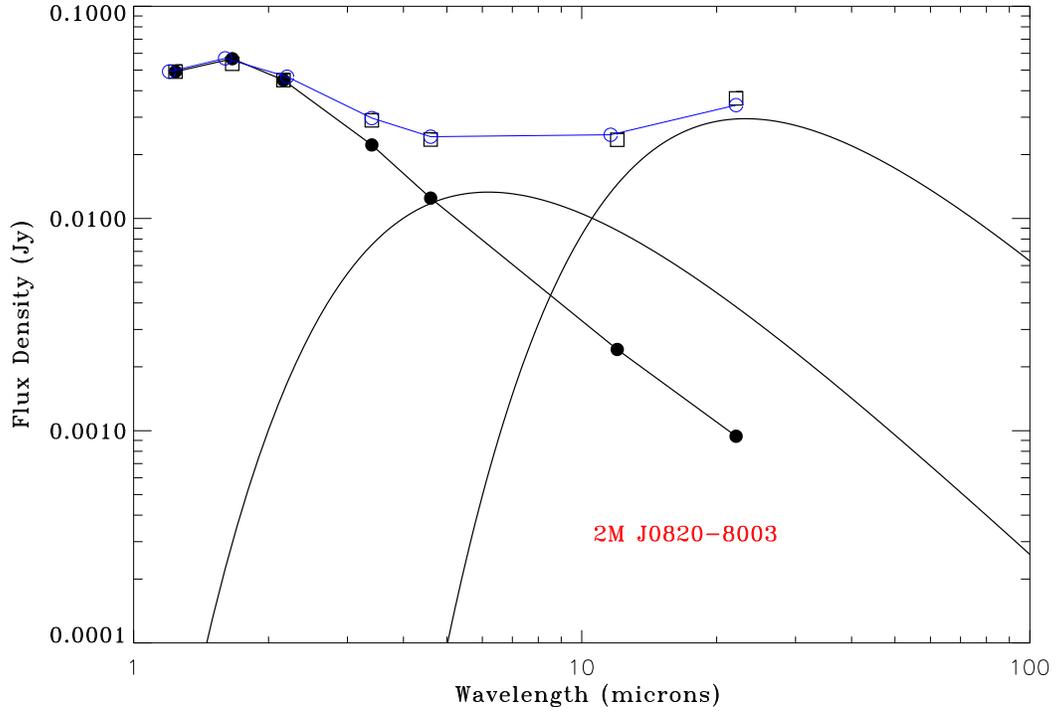}
\end{center}
\caption{Open squares are the J, H, and K (2MASS) fluxes of 2M
J0820-8003 and the {\it WISE} fluxes at 3.4, 4.6, 11.6, and 22.1
$\mu$m.  The filled black symbols are the fluxes in the same bands
of a field M4V star normalized to 2MASS J0820-8003 at J.  The
solid black lines are dilute black bodies at temperatures 220 and
825 K.  The solid blue line is the sum of the stellar SED and the
two black bodies. It represents a fit to the spectrum of the star 
and its disk.} 
\end{figure}

\end{document}